\documentclass[conference]{IEEEtran}
\IEEEoverridecommandlockouts
\usepackage{cite}
\usepackage{algorithm}
\usepackage{algpseudocode}
\usepackage{graphicx}
\usepackage{multirow}
\usepackage{hyperref}
\usepackage{float}
\usepackage{xcolor}
\usepackage{amsmath,amssymb,amsfonts}
\hypersetup{
    colorlinks=true,
    citecolor=black,
    linkcolor=black,
    filecolor=black,      
    urlcolor=black,
}

\renewcommand{\baselinestretch}{1.06}
    
\begin{document}

\title{Hunter in the Dark: \\
Discover Anomalous Network Activity\\ Using Deep Ensemble Network
}

\author{\IEEEauthorblockN{Shiyi Yang}
\IEEEauthorblockA{\textit{School of CSE} \\
\textit{University of New South Wales}\\
Sydney, Australia \\
z5223292@cse.unsw.edu.au}
% \and
% \IEEEauthorblockN{Nour Moustafa}
% \IEEEauthorblockA{\textit{School of EIT} \\
% \textit{University of New South Wales}\\
% Canberra, Australia \\ nour.moustafa@unsw.edu.au}
\and
\IEEEauthorblockN{Hui Guo}
\IEEEauthorblockA{\textit{School of CSE} \\
\textit{University of New South Wales}\\
Sydney, Australia\\
h.guo@unsw.edu.au}
\and
\IEEEauthorblockN{Nour Moustafa}
\IEEEauthorblockA{\textit{School of EIT} \\
\textit{University of New South Wales}\\
Canberra, Australia \\ nour.moustafa@unsw.edu.au}
}

\maketitle

\begin{abstract}

Machine learning (ML)-based intrusion detection systems (IDSs) play a critical role in discovering unknown threats in a large-scale cyberspace. They have been adopted as a mainstream hunting method in many organizations, such as financial institutes, manufacturing companies and government agencies. However, existing designs achieve a high threat detection performance at the cost of a large number of false alarms, leading to alert fatigue. To tackle this issue, in this paper, we propose a neural-network-based defense mechanism named DarkHunter. DarkHunter incorporates both supervised learning and unsupervised learning in the design. It uses a deep ensemble network (trained through supervised learning) to detect anomalous network activities and exploits an unsupervised learning-based scheme to trim off mis-detection results. For each detected threat, DarkHunter can trace to its source and present the threat in its original traffic format. Our evaluations, based on the UNSW-NB15 dataset, show that DarkHunter outperforms the existing ML-based IDSs and is able to achieve a high detection accuracy while keeping a low false positive rate.

\end{abstract}


\section{Introduction} \label{sec:introduction}

In network security, the battle between hackers and defenders is a never-ending game. Hackers launch crafted attacks by exploiting vulnerabilities in websites, loopholes in operating systems, flaws in applications and so on. Defenders, on the other hand, would often maintain the signatures of known cyber threats in order to identify corresponding intrusions and deploy security policies to prevent those malicious activities. The policies should ideally be secure and robust. But once the security policies are exposed, hackers can tweak their strategies to launch completely new attacks, which will, in turn, trigger defenders to generate new rules to counter the new attacks. This signature-based attack-and-counter-attack battle can go on and on, making safeguarding network security a very taxing endeavor. One effective mechanism to handle such a situation is leveraging the availability of massive network data and the intelligence of machine learning (ML) to develop an intrusion detection system (IDS) that can automatically learn attack behaviours from the network traffic and hence discover not only known but also unknown cyber attacks \cite{9346261}.

While ML-based IDSs have a high capability of novel threat perception, existing designs achieve a high attack detection performance often at the cost of a large number of false alarms. Traditional ML methods\cite{buczak2015survey}, due to their limited scalability to the large network traffic, treat most new and unanticipated behaviours as anomalies even if some of them are legitimate traffic\cite{9129472}.
The high level of false positive predictions would cause alert fatigue and likely make real threats miss in the noise of false alarms and fail to get timely attention from the security team \cite{9064976}.
The advanced deep learning (DL) approaches with deep neural networks (DNNs) can effectively mitigate the problem. DNN can learn features at various levels of abstraction, thus accomplishing much better generalization performance -- the adaptability to previously unseen data -- than the traditional ML \cite{boukhalfa2020lstm}. However, existing DL designs are still not mature enough\cite{wu2020pelican}, which leaves room for improved design solutions.

In this paper, we propose a new defense mechanism, DarkHunter, for network intrusion detection. DarkHunter comprises three modules: stream processor, detection engine and correlation analyzer. The stream processor is a data preprocessing module that converts network traffic flows into machine-learning-oriented data records. The detection engine is a specially designed deep ensemble neural network for attack detection.
The correlation analyzer is an analysis module that can filter out mis-predictions from the detection engine and generate user-friendly threat alerts for the security team.

Our main contributions are summarized as follows.

\begin{itemize}

    \item We develop a deep ensemble neural network, EnsembleNet, for efficient threat detection. Unlike the traditional ensemble designs, which are mainly based on simple and weak ML models, our ensemble design is constructed with the DNN models so that the high learning potential of DNN can be utilized for good detection performance. 
    
    \item We present a set of DNN designs as the sub classifiers for EnsembleNet. Each classifier is built with the CNN and RNN subnets that are connected in a way such that features learned by the subnets can be reused and hence the gradient vanishing and performance degradation problems in deep learning can be effectively mitigated. 
     
    \item To consolidate the prediction results from the sub classifiers in EnsembleNet, we propose a greedy majority voting algorithm for threat classification. The algorithm can be applied not only to binary classification but also to multi-class classification and is scalable for a large ensemble network with many sub classifiers.  

    \item To reduce the mis-predictions from EnsembleNet, we apply the principle component analysis and local outlier factor techniques to filter out some false alarms and find back some threats missed by EnsembleNet.
    
    \item For the detection results to be useful to the security team, we propose an alert-output enhancement design. The design restores the threats (detected by the neural network) to their human-understandable raw traffic format and produces alerts of the current threats in the order of their severity so that the security team can make prompt responses and hence maximally reduce the security risk. 

    \item With the above component-level designs, we develop the defense mechanism, DarkHunter. Our evaluations, on a near real-world dataset, UNSW-NB15 \cite{moustafa2015unsw}, show that DarkHunter outperforms existing ML-based IDSs and is able to achieve a high detection accuracy while keeping a low false alarm rate.

\end{itemize}

The remainder of this paper is organized as follows. In Section \ref{section_2}, we briefly review the existing designs for network intrusion detection. The defense mechanism, DarkHunter, is presented in Section \ref{section_4}. The experimental evaluation and discussion are given in Section \ref{section_5}. The paper is concluded in Section \ref{section_6}.

\section{Background and Related Work} \label{section_2}

An intrusion detection system can be a software application or a hardware device and is often deployed in a network to monitor in-and-out traffic and generate alerts of any suspicious activities or policy violations to the security team for threat identification and elimination \cite{wu2020pelican}. Two kinds of IDSs that are currently used in the industry are rule-based and anomaly-based (aka ML-based). They are briefly discussed in the following two sub-sections.

\subsection{Rule-Based Pattern Matching Intrusion Detection System}

A rule-based IDS discovers threats by matching attack signatures or patterns against a pre-defined blacklist, which is effective to identify known attacks. Due to its stability and dependability, the IDS is by far the most widely used in real-world business environments. Snort\footnote{Snort: https://www.snort.org}, Suricata\footnote{Suricata: https://suricata-ids.org} and Zeek\footnote{Zeek: https://www.zeek.org} are representative security products. These tools monitor traffic streams, especially checking for some specific features, such as a certain protocol or suspicious IP addresses or a byte pattern existing in packet payloads like some URI or USER-AGENT that may indicate some malicious activities. Once these features match their well-defined rules, an alarm is triggered. Zeek is more flexible than the other two in that it adds a programmatic interface, allowing users to customize traffic analysis according to different network environments. In recent years, with the increasing complexity and number of unknown cyber attacks, such as distributed denial-of-service (DDoS) attacks and advanced persistent threats (APTs) \cite{ashraf2021iotbot}, the weakness of rule-based IDS has gradually been exposed. Hand-designed attack signatures contain excessively detailed descriptions of known attacks, making it almost impossible for such an IDS to discover novel threats. 
Another disadvantage of the rule-based IDS is that it requires the security team to frequently and manually update the signature and rule database, which is tedious and time consuming. To tackle such issues, the anomaly-based IDS comes into play.

\subsection{Data-Centric Anomaly-Based Intrusion Detection System}

The anomaly-based IDS leverages the available massive learning data and the heuristics generated from machine learning to create a model of trustworthy network activities; Any activities deviated from the model can be treated as threats, hence making it possible to detect new attacks. Two typical machine learning methods can be used in building such a detection model \cite{laskov2005learning}. A brief description is given below.

\subsubsection{Unsupervised Machine Learning Methods}

Unsupervised learning focuses on finding patterns, structures or knowledge from unlabeled data. Clustering and anomaly detection are two main related techniques used in the early anomaly-based IDSs \cite{laskov2005learning}, and K-Means \cite{jianliang2009application} and local outlier factor (LOF) \cite{alshawabkeh2010accelerating} are the most representative algorithms of the two techniques. K-Means divides the given data into K homogeneous and well-separated clusters, and each network traffic record belongs to the cluster with the nearest mean. In contrast, LOF finds anomalous traffic records by calculating the local density deviation of the records with respect to their neighbors. Compared to K-Means, LOF is more efficient for attack detection and widely used.

The main advantage of unsupervised learning is that it avoids the costly and time-consuming data labeling process, hence freeing up human and computing resources for other important tasks, such as counter-attack. However, lack of labeling information makes it hard to interpret the learning results \cite{cs259d}.

\subsubsection{Supervised Machine Learning Approaches}

Supervised learning constructs a predictable profile with a set of well-tagged network traffic records. Compared to unsupervised learning, it is more suitable to practical implementations \cite{suaboot2020taxonomy}. The supervised learning based designs can be basically divided into two groups: classical ML-based and advanced DL-based.

For network intrusion detection, the classical ML methods \cite{buczak2015survey} can be further considered as two kinds: individual classifiers and ensemble classifiers. Among many individual classifiers, kernel-based support vector machine (SVM) \cite{krishnaveni2020anomaly} and probability-based naive bayes (NB) \cite{zhang2018network} are two effective designs. SVM uses a kernel trick to implicitly map the input data to high-dimensional feature space for efficiently classifying security data, whereas NB makes classification by applying Bayes' Theorem and assuming that features in the data are independent.
In contrast, ensemble classifiers, such as random forest (RF) \cite{9155656} and adaptive boosting (AdaBoost) \cite{shahraki2020boosting}, integrate several weak learners into a stronger learner to improve the generalizability and robustness of the classifier. And the final learning result from the classifier is generated through an integration scheme. For example, RF uses Bagging algorithm and AdaBoost employs Boosting algorithm.

As the scale and complexity of network traffic increases in recent years, the traditional ML techniques have suffered from the performance optimization bottleneck, due to so-called ``the curse of dimensionality" issue \cite{bengio2006curse}. Consequently, even though the classical ML-based design is able to discover new attacks, it achieves high detection rates at the cost of high false alarms, resulting in alert fatigue.

Deep learning offers a promising solution to the above problem. DL is based on a neural network with a stack of learning layers. It can self-learn features from dataset and can generate new data representations from the previous learning results. Multilayer perceptron (MLP) \cite{9129472} is an early kind of feed-forward artificial neural network (ANN) with multiple layers and non-linear activation functions. It has a fully connected structure, which incurs a large amount of computation and hence restricts its learning efficiency on heavy network traffic. Convolutional neural network (CNN) \cite{9064976} and recurrent neural network (RNN) \cite{boukhalfa2020lstm} are two modern paradigms that apply parameter sharing techniques to reduce computational costs and are more suitable to network intrusion detection.

\begin{figure*}[t]
    \centering
    \includegraphics[width=0.98\linewidth]{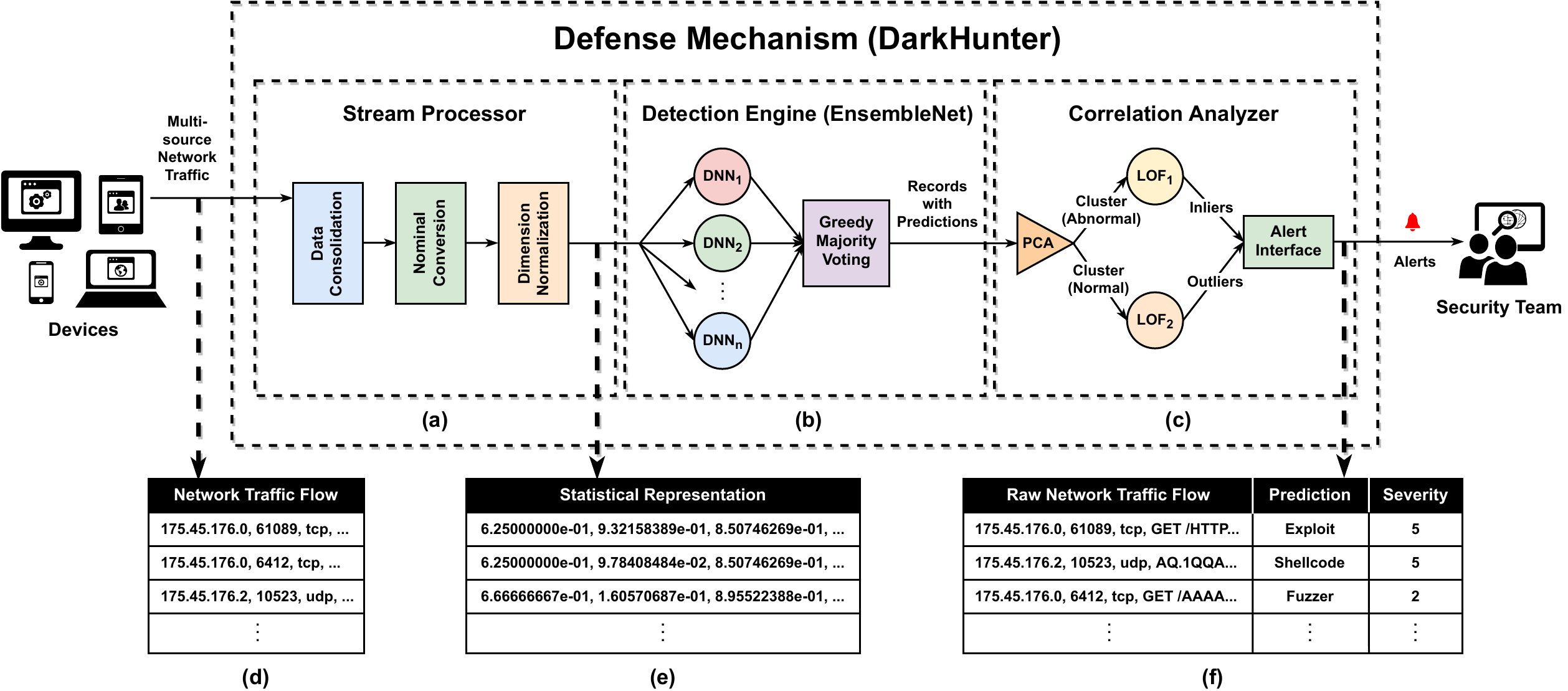}
    \caption{DarkHunter System Overview}
    \label{fig:system_overview}
\end{figure*}

CNN extracts the spatial-oriented features through convolution calculation from the learning dataset, whereas RNN establishes loop connections to capture the temporal-oriented features in the dataset. The common designs in the CNN family are primitive CNN (ConvNet) \cite{9064976} and depthwise separable CNN (DSC) \cite{chollet2017xception}, where DSC has a much lower computational cost due to the reduced multiplications. In the RNN family, long short-term memory (LSTM) \cite{boukhalfa2020lstm} solves the gradient vanishing problem and the inability of acquiring long-term dependencies in the vanilla RNN and hence is widely used in many application areas. Gated recurrent unit (GRU) \cite{cho2014learning}, on the other hand, is a simplified LSTM with a fewer number of gates and has a much lower computational cost.

Intuitively, to learn the features at various levels of abstraction to improve the generalization capability of the neural network model, the neural network is required to be deeper. However, 
as the depth of the network increases, the network will suffer from performance degradation problem \cite{wu2020pelican}. Densely-ResNet \cite{9343142} and DualNet \cite{9346261} are two state-of-the-art designs that address this issue by reusing features. Densely-ResNet is a densely connected residual network that can be expanded in both depth and width dimensions. 
But with the increase in width, the computational cost of the model increases greatly.
DualNet, by comparison, is a two-stage neural network architecture, where the first stage aims to maximally capture spatial and temporal features and the second stage improves the detection efficiency by targeting important features for the final learning outcome.

Although existing DL-based IDSs have demonstrated better generalization performance than traditional ML-based IDSs, the anomaly-based designs are largely at the research investigation stage. There is still room for improvements, especially in terms of the detection accuracy and interpret-ability of the detection results. As such, we put forward a new design, which is presented in the next section.

\section{DarkHunter} \label{section_4}

DarkHunter takes the synergy of supervised learning and unsupervised learning. It consists of three modules: stream processor, detection engine and correlation analyzer, as shown in Fig.~\ref{fig:system_overview}. The design of each module is explained below.

\subsection{Stream Processor} \label{section_3_a}

The stream processor is for data preprocessing. It converts network traffic flows, as illustrated in Fig.~\ref{fig:system_overview}(d) into the statistical data records suitable for neural network learning, as depicted in Fig.~\ref{fig:system_overview}(e). The module contains three functions, as shown in Fig.~\ref{fig:system_overview}(a). Each of them is elaborated below.

% Network traffic flows contain many overly specific features, such as IP addresses and protocols, which will result in poor generalization performance when these features are used directly on ML \cite{9343142}. 

% To solve the problem, we introduce a stream processor as a data entrance of DarkHunter. As illustrated in Fig. \ref{fig:system_overview}(a), it is made up of three data pre-processing functions. The module converts multi-source traffic flows, as shown in Fig. \ref{fig:system_overview}(d), into their statistical representatives involving latent patterns of legitimates or anomalies that can be effectively learned and detected by the detection engine, as demonstrated in Fig. \ref{fig:system_overview}(e). Three functions are elaborated below.

\begin{enumerate}

\item \textit{Data Consolidation.} The traffic data can be collected from multiple sources and have various formats, such as .argus, .log, and .json. The data consolidation merges the data records with a unified format. In addition, to ensure the data quality, the function also removes duplicate records and replaces missing data with the mean value of the related features.

\item \textit{Nominal Conversion.} There are many categorical features in the traffic flow, such as IP address and protocol, which cannot be fed straight into the neural network. The nominal conversion applies label encoding technique \cite{hancock2020survey} to convert the textual notation into a machine-readable form. Since the function employs digital codes to represent long textual values, the amount of computation by the neural network is reduced.

\item \textit{Dimension Normalization.} The features with larger magnitudes in the traffic data records may dominate in model fitting, leading to biased predictions. The dimension normalization uses the min-max normalization \cite{garcia2015data} to reconstruct data in each feature on a scale of 0 to 1 so that all features contribute to model fitting equally. The normalization also improves the learning stability and accelerates the back-propagation in training.

\end{enumerate}

\subsection{Detection Engine (EnsembleNet)}

The detection engine is an ensemble neural network, as shown in Fig.~\ref{fig:system_overview}(b). Unlike the traditional ensemble designs (such as RF), which are constructed with weak ML models, EnsembleNet in DarkHunter aims for high detection performance and hence uses the strong DNN models.

\begin{figure}[t]
    \centering
    \includegraphics[width=0.98\linewidth]{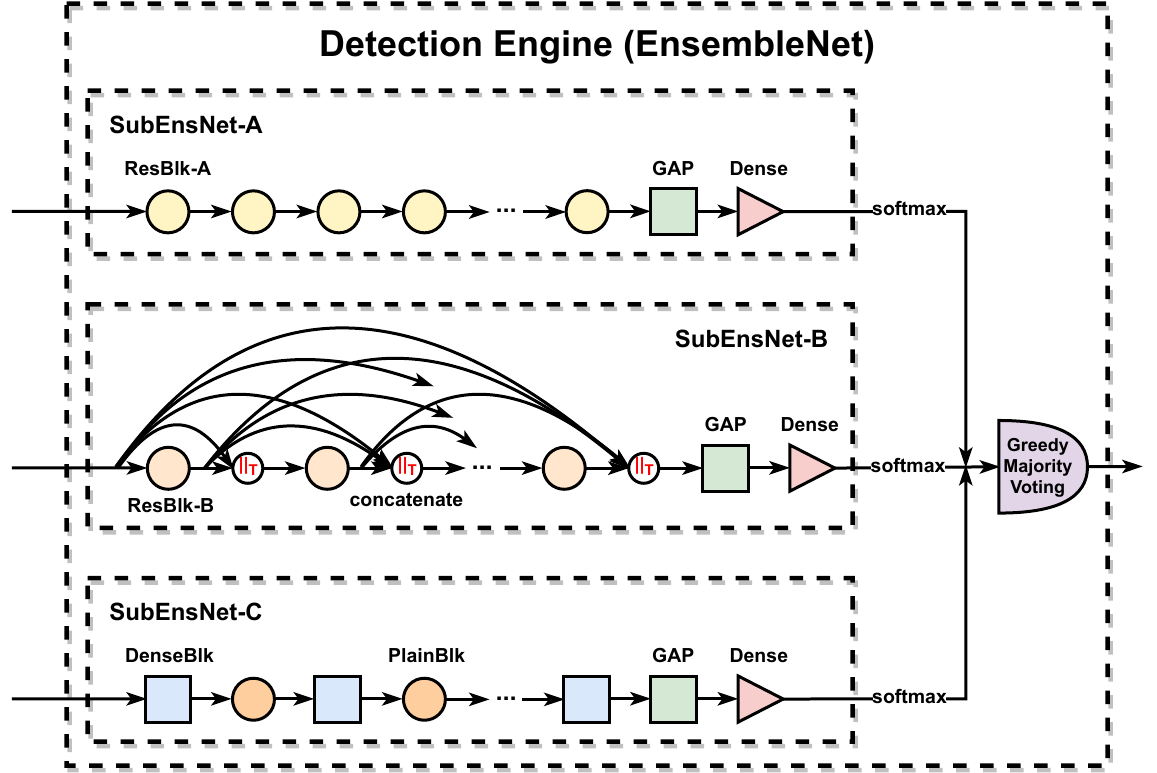}
    \caption{Detection Engine of DarkHunter}
    \label{fig:detection_engine}
\end{figure}

Fig.~\ref{fig:detection_engine} illustrates the overall architecture of an EnsembleNet with three DNNs.  For high efficiency, we want those DNN to have the following attributes: \textit{1) capable of spatial-temporal learning, 2) able to reuse learning features, and 3) having a low computational cost.} To this end, we build the DNNs on specially-designed blocks: plain blocks (PlainBlks), residual blocks (ResBlks) and dense blocks (DenseBlks). Each has a different complexity. In fact, PlainBlk is a building block of ResBlks and DenseBlks. Their designs are discussed below.

\subsubsection{Plain Block (PlainBlk)}

The plain block design is based on our previous work\cite{9346261}. It contains a CNN for spatial feature extraction and a RNN for temporal feature extraction. To reduce the computational cost of the block, we adopt the simplified versions of CNN and RNN, namely DSC \cite{chollet2017xception} and GRU \cite{cho2014learning}, as has been explained in Section \ref{section_2}. Here, DSC uses rectified linear unit (ReLU) as the activation function, whereas GRU uses both sigmoid and hyperbolic tangent (Tanh) as the activation function and sigmoid is applied to the recurrent step.

\begin{figure}[t]
    \centering
    \includegraphics[width=0.98\linewidth]{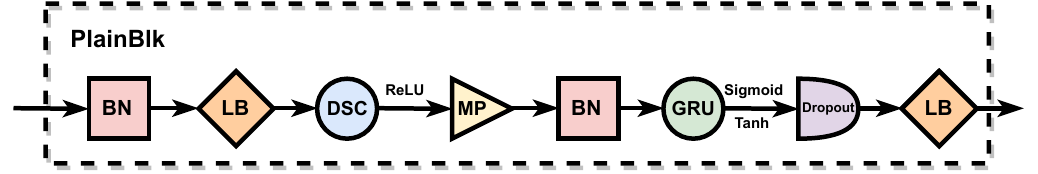}
    \caption{Plain Block}
    \label{fig:plain_blk}
\end{figure}

Fig. \ref{fig:plain_blk} shows the structure of PlainBlk. As can be seen from the figure, apart from the DSC and GRU subnets, we also introduce batch normalization (BN) \cite{ioffe2015batch} to standardize each mini-batch in training to reduce the internal covariate shift, thus accelerating DNN fitting and decreasing the final generalization errors. In addition, there are a max-pooling (MP) layer after DSC and a regularizer dropout \cite{srivastava2014dropout} after GRU; This is to prevent overfitting and further reduce the computational cost. We also add a linear bridging (LB) layer to transform non-linear parameter layers into a linear space, which is helpful for stabilizing the learning process.

Intuitively, a deeper neural network formed by stacking more PlainBlks can achieve better detection performance. However, as the network goes deeper, the network would suffer performance degradation \cite{wu2020pelican}. The main reason is that as the network depth increases, the learned features gradually become extremely specific but far away from their original meanings, eventually resulting in gradient vanishing \cite{9343142}. One of the effective methods to solve this optimization obstacle is \textit{feature reuse}. Feature reuse keeps the features initially learned from the shallow parameter layers and makes them available at the deep layers to retain the originality of features.
This is done by connecting shallow layers to deep layers, and combining the early learned features with the later learned features. The combination can be in two operation modes: \textit{add} and \textit{concatenation}, and the concatenation operation can be performed on the data records along either the feature-oriented dimension or the timestep-oriented dimension, which leads to our other block designs: ResBlk and DenseBlk.

\subsubsection{Residual Block (ResBlk)}

\begin{figure}[b]
    \centering
    \includegraphics[width=.98\linewidth]{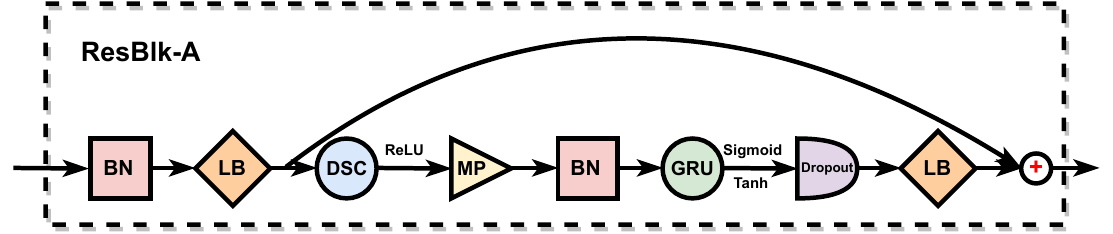}
    \caption{Residual Block Type A}
    \label{fig:res_blk_a}
\end{figure}

\begin{figure}[t]
    \centering
    \includegraphics[width=.98\linewidth]{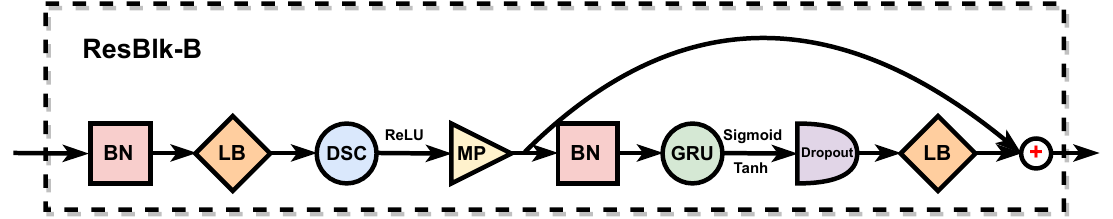}
    \caption{Residual Block Type B}
    \label{fig:res_blk_b}
\end{figure}

ResBlk incorporates the residual learning \cite{he2016deep} into PlainBlk to handle performance degradation. It adds a shortcut connection and uses ``add" to combine the features from both connected layers. ResBlk has two versions of design: ResBlk-A and ResBlk-B.  Fig.~\ref{fig:res_blk_a} shows the structure of ResBlk-A, where the input of DSC is connected to the output of the last layer. While for ResBlk-B, there is a slight difference in the shallow layer to be connected, as shown in Fig.~\ref{fig:res_blk_b}, which will be further discussed later.
The shortcut facilitates the forward propagation of activations and the backward propagation of errors, thus avoiding gradient vanishing. It is worth noting that the summation operation requires that the tensors to be added have the same shape.

\begin{figure}[b]
    \centering
    \includegraphics[width=.98\linewidth]{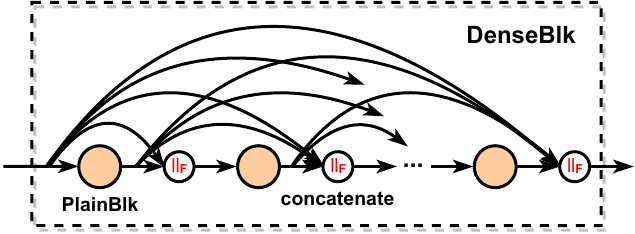}
    \caption{Dense Block}
    \label{fig:dense_block}
\end{figure}

\subsubsection{Dense Block (DenseBlk)}

DenseBlk is formed with a set of densely-connected PlainBlks, as shown in Fig.~\ref{fig:dense_block}, where each PlainBlk receives a concatenation of the input and the data from all its preceding PlainBlks. In such a dense connectivity pattern\footnote{The idea of building dense connections is originated from densely connected convolutional networks \cite{huang2017densely}, which have shown good performance for object recognition tasks.}, features at various levels of abstraction can be fully learned. Moreover, the flow of gradients within the network can be significantly strengthened, thus addressing the performance degradation caused by the vanishing gradient. 
For DenseBlk, the concatenation along the feature-oriented dimension is used, as denoted by symbol $\vert\vert_F$ in Fig.~\ref{fig:dense_block}. \\

Based on the above basic building blocks, we develop three DNN sub-classifiers for EnsembleNet.\\

\subsubsection{Sub-classifiers of EnsembleNet}

The three sub-classifiers named as SubEnsNet-A, SubEnsNet-B and SubEnsNet-C, as shown in Fig.~\ref{fig:detection_engine}, are presented below.

\textit{SubEnsNet-A.} SubEnsNet-A is a deep residual neural network. It is constructed with a series of residual blocks, ResBlk-A blocks, followed by a global average pooling (GAP) layer and a dense layer with the softmax activation function. The GAP layer is to further strengthen corresponding relationships
between features and the classification categories and the dense layer is to determine the final detection result. As can be seen from Fig.~\ref{fig:detection_engine}, both GAP layer and dense layer are also applied to other two sub-classifiers (for the same purposes).

\textit{SubEnsNet-B.} SubEnsNet-B is a deep densely connected residual neural network, which is also built upon a group of ResBlks but with a dense connection structure using the concatenation along the timestep-oriented dimension (denoted as $\vert\vert_T$), as shown in Fig.~\ref{fig:detection_engine}. In addition, instead of using ResBlk-A, ResBlk-B is used. The reason is that the data dimension from the concatenation would increase, making the shape of tensors to be added inconsistent if ResBlk-A was adopted. To handle the problem, we leverage the down-sampling ability of the MP and move the shortcut just after the MP (see Fig.~\ref{fig:res_blk_b}) so that the tensor shape can be adjusted while the local originality of features can be mostly retained.

\textit{SubEnsNet-C.} SubEnsNet-C is a deep dense neural network that is different from the first two. As can be seen from Fig.~\ref{fig:detection_engine}, SubEnsNet-C is established with DenseBlks and PlainBlks in an interleaved arrangement pattern. Since PlainBlk is densely connected along the feature-oriented dimension in DenseBlk, DenseBlk is vulnerable to the curse of dimensionality problem \cite{bengio2006curse}. This is because the dimensionality of feature space tends to grow exponentially when stacking more PlainBlks in one DenseBlk. As a result, the prediction ability of the model will decrease significantly. To solve the issue, we interleave DenseBlk with PlainBlk to leverage the internal down-sampling capability of PlainBlk to limit the growth of feature space while maintaining both spatial and temporal features during the dimensionality reduction \cite{9346261}. In this way, we can build a very deep SubEnsNet-C.\\

In summary, we have built three extensible DNNs that allow themselves to go deeper for a good intrusion detection performance. When we put them into EnsembleNet, even better performance can be achieved, which is closely related to how to combine the prediction results from these sub-classifiers. Our design is given below. \\

\subsubsection{Greedy Majority Voting Algorithm} 

There are some existing aggregation algorithms for ensemble classifiers, such as Bagging \cite{9155656} and Boosting \cite{shahraki2020boosting}, as mentioned in Section~\ref{section_2}. Boosting is a sequential process on a set of sequentially-performed sub-classifiers, which could result in considerable computing time if these classifiers were DNNs.
Bagging, on the other hand, can be performed on the parallel sub-classifiers. But each of the classifiers only works on a small subset of the input data, which may cause model underfitting if DNN was used. In addition, in Bagging, random selection is used when the sub-classifiers come to multiple tied-voting results, which is not effective for multi-class classification tasks, and binary classification tasks based on an even number of voters. Here, we propose a different integration algorithm, Greedy Majority Voting, for EnsembleNet.

Our algorithm combines the idea of majority voting with the
detection performance of each DNN sub-classifier, where each DNN learns the entire dataset and is trained and tested in parallel.
Unlike normal majority voting, which is largely for binary classification and often assumes odd number of voters, our greedy majority voting algorithm also supports even number of classifiers and can handle both binary and multi-class classifications. 
Considering that for a DNN model, Accuracy indicates its generalization capability and Precision indicates whether a high threat detection capability of the model comes at the cost of high false alarms, we use the score obtained from these two metrics to represent the DNN performance in our algorithm. The algorithm is explained below. 

\begin{figure}[t]
    \centering
    \includegraphics[width=.98\linewidth]{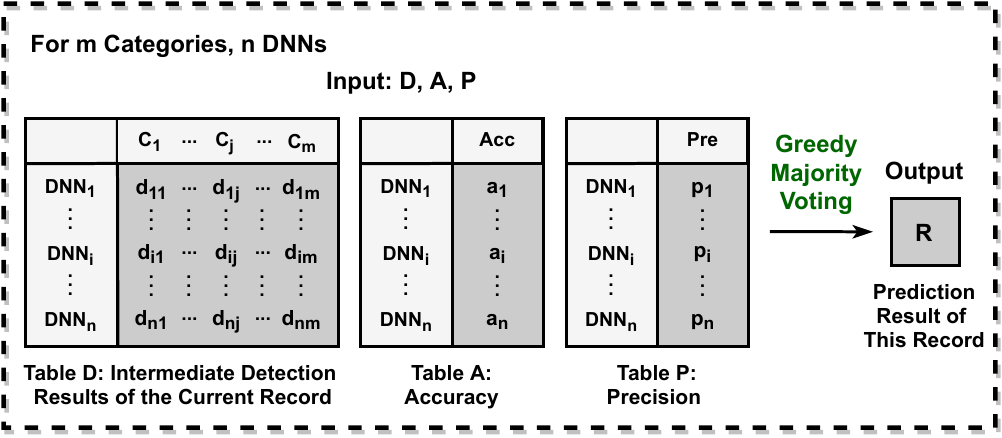}
    \caption{The Input and Output of Greedy Majority Voting Algorithm}
    \label{fig:majority_voting}
\end{figure}

Assume there are $m$ detection categories, $C_1$, ..., $C_m$ and $n$ DNNs, $DNN_1$, ..., $DNN_n$. We use three tables to present the information available to the algorithm, as shown in Fig.~\ref{fig:majority_voting}. Table $D$ holds the detection results from the $n$ DNNs for the current network traffic record to be classified, where $d_{ij}$ is the decision of $DNN_i$ on whether the traffic record belongs to category $C_j$; if yes, $d_{ij}$ is 1, otherwise 0. For a traffic record, each DNN will predict one and only one category to be true, therefore, the sum of each row in Table $D$ is 1. Table $A$ stores the accuracy of each DNN, where $a_i$ is the accuracy of $DNN_i$. The precision of the DNNs is saved in Table $P$; on the same notion, $p_i$ is the precision of $DNN_i$.  The result returned from the algorithm is saved in $R$ that holds the final prediction of the ensemble network.

% \begin{algorithm}[t]
% \caption{Greedy Majority Voting Algorithm} \label{alg:greedy}

% \hspace*{\algorithmicindent} \textbf{Input:} $D$, $A$, $P$\\
% \hspace*{\algorithmicindent} \textbf{Output:} $R$

% \begin{algorithmic}[1]
% \For{\textbf{each} $j=1, 2, ..., m$}
% \State $v_j = $ Sum($d_{1j} + d_{2j} + ... + d_{nj}$)
% \EndFor
% \State $S = $ Max(\{$v_j$, $j=1, 2, ..., m$\})

% \If {$|S|= 1$} 
% \Comment{Only one element in $S$}
% \State $R = C_j$, $v_j \in S$
% \Else 

% \State $S =$ Max(\{$a_i$, for all $i$, where $d_{ij} = 1$ and $v_j \in S$\})

% \If {$|S|= 1$} 

% \State $R = C_j$, where $d_{ij} = 1$ and $a_i \in S$ 

% \Else

% \State $S =$ Max(\{$p_i$, for all $i$, where $a_i \in S$\})

% \If {$|S|= 1$}

% \State $R = C_j$, where $d_{ij} = 1$ and $p_i \in S$
% \Else

% \State $R = C_j$, where $d_{ij} = 1$ and $p_i$ is the first element of $S$

% \EndIf
% \EndIf
% \EndIf
% \State \textbf{return} $R$

% \end{algorithmic}

% \end{algorithm}

\begin{algorithm}[t]
\caption{Greedy Majority Voting Algorithm} \label{alg:greedy}
\hspace*{\algorithmicindent} \textbf{Input:} $D$, $A$, $P$\\
\hspace*{\algorithmicindent} \textbf{Output:} $R$
\begin{algorithmic}[1]
\For{\textbf{each} j = 1, 2, ..., m}
\State $v_j = $ Sum($d_{1j} + d_{2j} + ... + d_{nj}$)
\EndFor
\State $V = $ Max(\{$v_j$, j = 1, 2, ..., m$\})$
\State $S = $ getPredictionWithMaxVote($V$)
\If {$|S|= 1$} 
\Comment{Only one element in $S$}
\State $R = C_j$, $C_j \in S$
\Else
\State $S = $ getPredictionWithHighAccuracy($S, A$)
\If {$|S|= 1$}
\State $R = C_j$, $C_j \in S$
\Else
\State $S =$ getPredictionWithHighPrecision($S, P$)
\If {$|S|= 1$}
\State $R = C_j$, $C_j\in S$
\Else
\State $R = $ getFirstPrediction($S$)
\EndIf
\EndIf
\EndIf
\State \textbf{return} $R$
\end{algorithmic}
\end{algorithm}

Algorithm~\ref{alg:greedy} describes steps of the greedy majority voting. Given the predictions from the DNNs, $D$, the DNNs' accuracy in $A$ and precision in $P$, the algorithm first collects votes for each prediction category (line 1 - line 3). Then it selects the predictions with the highest votes and saves it in $S$ (line 4 - line 5). If there is only prediction in $S$, it returns the prediction and the algorithm stops (line 6 - line 7); Otherwise, if there is more than one prediction in $S$ (e.g. two predictions are tied in votes), it removes those predictions from $S$ that are generated by the DNNs with lower performance, which is based first on their accuracy and then on their precision if required (line 8 - line 15). After that, if there is still more than one prediction in $S$, return the first one (line 16 - line 17).

The algorithm produces the prediction for each traffic record. The predictions from EnsembleNet may not always be accurate, which can be improved by our next design module, as given below.

\subsection{Correlation Analyzer}

The analyzer aims to enhance the detection performance and the understand-ability of the detection result, which is achieved by analysing two kinds of correlations: correlation between traffic records and correlation between the neural-network prediction output and the raw traffic stream.

\subsubsection{Enhancing Detection Performance}

As stated in \cite{kill_chain} when the cyber kill chain framework is investigated, there are seven steps that adversaries must complete to reach their goals. A malicious traffic flow often belongs to one of the steps (e.g. Reconnaissance flow belongs to Step 1 and Exploit flow belongs to Step 4). Therefore, \textit{attack-related flows have high correlations with themselves but low correlations with benign flows, and vice versa}.

We put the detected anomalous and normal network records from EnsembleNet into two clusters: Abnormal and Normal. The mis-predictions in each cluster are the outliers of the cluster. So we can trim off some false alarms from the Abnormal cluster and find back some attacks from the Normal cluster, for which we utilize two unsupervised learning techniques: principal component analysis (PCA)  \cite{garcia2015data} and local outlier factor (LOF) \cite{alshawabkeh2010accelerating}.

\begin{figure}[t]
    \centering
    \includegraphics[width=0.98\linewidth]{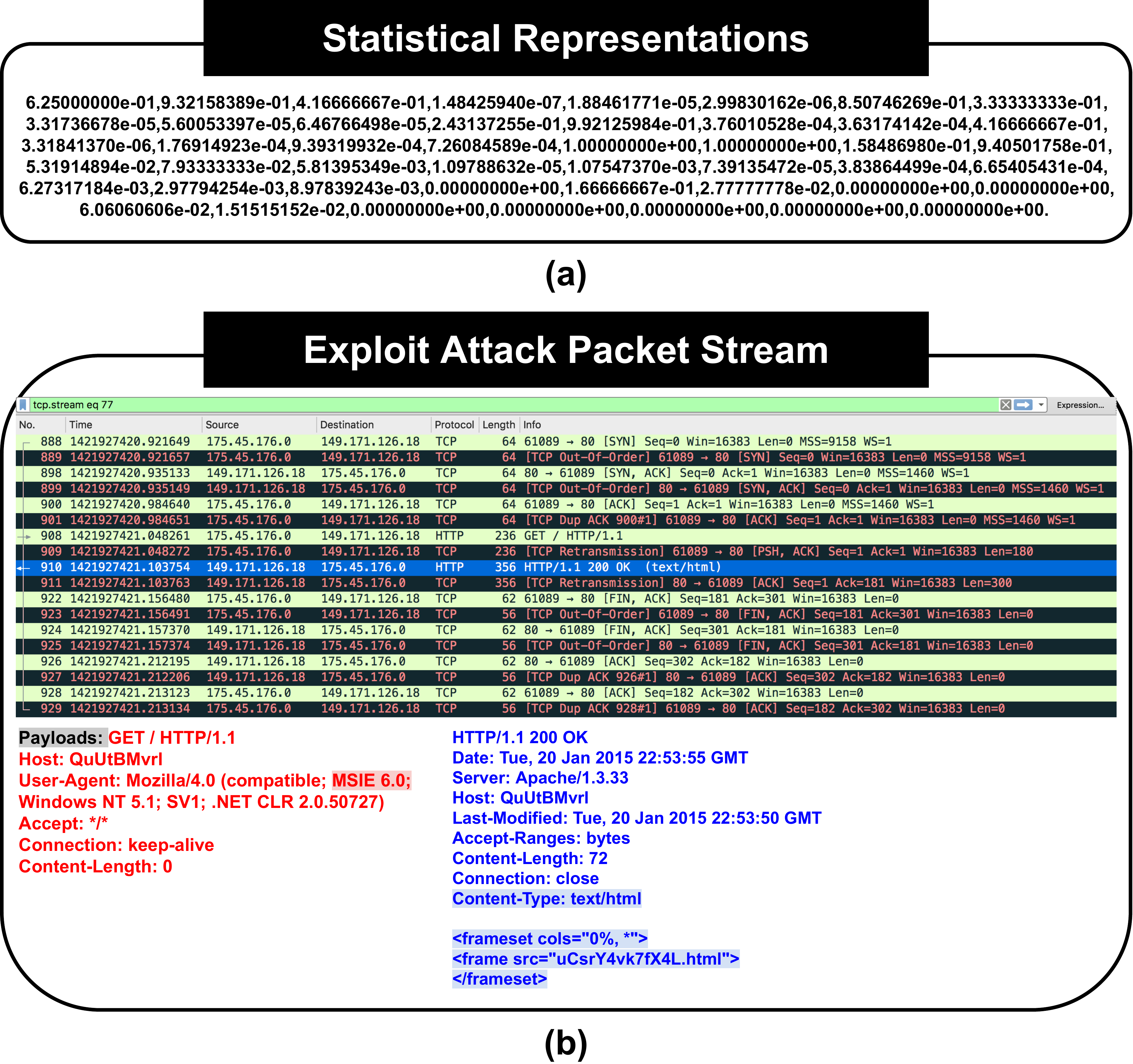}
    \caption{An Exploit attack within a TCP stream: Microsoft Internet Explorer Frameset Memory Corruption and the attack reference is CVE-2006-3637}
    \label{fig:interpretability}
\end{figure}

PCA can obtain dominant but eliminate redundant components from features of network traffic. We apply PCA to heighten the correlation between attack records, the correlation between normal records, and the difference between attack records and normal records, making it easier to identify outliers in each cluster by using LOF. LOF has been briefly described in Section \ref{section_2}. We leverage LOFs to integrate highly-correlated attack records from the Abnormal cluster into a sub-cluster and highly-correlated normal records from the Normal cluster into another sub-cluster. Hence, records that are inside the clusters but outside the sub-clusters can be considered as outliers (i.e. mis-predictions). As a result, the inliers from the Abnormal cluster and outliers from the Normal cluster form the final detection result from the neural network.

\subsubsection{Enhancing Detection Output} 

We want the threat alerts generated from the detection system to be understandable and useful to human security analysts. Therefore, we include a user-friendly alert interface in DarkHunter. The interface offers the following two functions.

\textit{Original Packet Stream Backtracking.} As shown in Fig. \ref{fig:system_overview}(e), the features used for learning are vectorized, making it difficult for security analysts to understand, interpret and analyze. 
Our investigation on the network traffic data shows that the network traffic streams can be uniquely identified by their source IP address, source port number, destination IP address, destination port number, flow start time and end time, and protocol.
We use this combined information to correlate a threat detected by the neural network to the related raw network flow in different source traces and restore the threat from the neural-network-used format to the human-understandable textual traffic format.

We adopt Wireshark\footnote{Wireshark: https://www.wireshark.org} to visualize the original packet streams. Fig.~\ref{fig:interpretability} shows an example for a detected Exploit attack. Its neural-network-used format is shown in Fig.~\ref{fig:interpretability}(a) and  Fig.~\ref{fig:interpretability}(b) is its restored raw network stream.

As can be seen from the figure, the stream contains 3-way handshake packets for establishing TCP connections, HTTP request packets, HTTP response packets and 4-way handshake packets for tearing down TCP connections. The payloads of the stream provide more valuable attack-related information. Traffic from the client to the server is colored in red, while traffic from the server to the client is colored blue. The payloads indicate \textit{`Microsoft Internet Explorer Frameset Memory Corruption \cite{exploit_attack}' (CVE-2006-3637 \cite{cve_2006_3637}):} a flaw in Microsoft Internet Explorer (MSIE) 6.0 (marked in pink) that makes the browser unable to properly handle various combinations of HTML layout components. The hacker exploits the vulnerability when rendering HTML using a crafted frameset (marked in blue), which results in memory corruption. 

By obtaining the raw traffic format, we can find more cyber threat intelligence and specific attack behavior from the payloads to get insight into the threat. More examples are demonstrated in Section \ref{section_5}.

\renewcommand{\baselinestretch}{1.1}
\begin{table*}[t]
\caption{Testing Performance of DarkHunter and Its DNN Components on UNSW-NB15 Testbed}
\begin{center}
\begin{tabular}{c||c|c|c|c|c|c|c}
\hline
\multirow{2}{*}{\textbf{Design}} & \multicolumn{2}{c|}{\textbf{ACC \%}} & \multirow{2}{*}{\textbf{DR \%}} & \multirow{2}{*}{\textbf{FAR \%}} & \multirow{2}{*}{\textbf{Precision \%}} & \multirow{2}{*}{\textbf{$F_1$ Score}} & \multirow{2}{*}{\textbf{Time/s}}\\
\cline{2-3} & \textbf{Multi-class Classification} & \textbf{Binary Classification} & & & & & \\
\hline
\hline
SubEnsNet-A & 75.45 & 87.53 & 97.57 & 24.76 & 82.84 & 89.60 & 631.17\\
\hline
SubEnsNet-B  & 75.71 & 87.28 & 97.56 & 25.32 & 82.52 & 89.41 & 718.83\\
\hline
SubEnsNet-C & 76.26 & 88.38 & 97.67 & 23.01 & 83.87 & 90.25 & 711.86\\
\hline
\textbf{EnsembleNet} & \textbf{77.27} & \textbf{88.66} & \textbf{98.12} & \textbf{22.92} & \textbf{83.99} & \textbf{90.51} & \textbf{718.88}\\
\hline
\textbf{DarkHunter} & \textbf{80.18} & \textbf{90.81} & \textbf{98.13} & \textbf{17.71} & \textbf{86.58} & \textbf{91.99} & \textbf{720.86}\\
\hline
\end{tabular}
\label{tab:ensemble_nets}
\end{center}
\vspace{-1em}
\end{table*}
\renewcommand{\baselinestretch}{1}

\textit{Threat Response Priority Assessment.} 
Different attacks may impose different levels of security risk.
Based on the recommendation of the security analysts in the industry, we group all cyber threats into five severity levels, ranging from level 1 (of the lowest) to level 5 (of the highest): \textit{level 1---low impact, level 2---possible impact, level 3---medium impact, level 4---significant impact and level 5---high impact.} 
For example, Analysis attack belongs to level 1; Reconnaissance
and Fuzzer attacks belong to level 2; DoS attack belongs to
level 3; Generic, Backdoor and Worm attacks belong to level
4; and Exploit and Shellcode attacks belong to level 5.
Based on our observations, the majority of false positives produced by the ML-based IDSs are the normal network traffic mis-classified as low-severity threats (e.g. Analysis). This may be due to their similar behaviors.

We output alerts of the attacks in the order of their severity, as demonstrated in Fig.~\ref{fig:system_overview}(f). so that the severest attacks can get immediate attention from the security team and be contained.

\section{Evaluation and Discussion} \label{section_5}

\subsection{Experimental Environment Settings}

Our evaluation is based on a cloud AI platform configured with an NVIDIA Tesla K80 GPU and a total of 12 GB of RAM. DarkHunter and other related IDS designs, to be used for comparison, are modeled in Python, with Tensorflow as the backend and the APIs of Keras libraries and scikit-learn packages.

\subsection{Dataset Selection}

We use UNSW-NB15 \cite{moustafa2015unsw} as a cyber threat assessment testbed. UNSW-NB15 was generated by IXIA PerfectStorm tool \cite{ixia}. The tool simulates a real-word network environment with millions of up-to-date attack scenarios that are updated regularly from the Common Vulnerabilities and Exposures (CVE) site \cite{cve} and the normal traffic. 
This dataset offers both real and synthesized anomalous network activities and covers nine typical attack types: Generic, Exploit, Fuzzer, Reconnaissance (Recon), DoS, Shellcode, Backdoor, Analysis and Worm. 
The dataset contains 2,540,044 traffic records with 47 features, where ten percent of traffic records were randomly selected for evaluation. To ensure the effectiveness of evaluation, we choose 70\% of selected records in random as a training set and the remainder as a testing set, where the testing samples are not used for training so that they can be applied to verify the generalization performance of detection models.

\subsection{Configuration of DarkHunter}

For detection engine (EnsembleNet) of DarkHunter, we configure SubEnsNet-A with 10 ResBlk-A blocks, SubEnsNet-B with 10 ResBlk-B blocks and SubEnsNet-C with two DenseBlks interleaved by one PlainBlk, where each DenseBlk has five PlainBlks. 
Therefore, SubEnsNet-A has 83 layers including 53 parameter layers; Similarly, SubEnsNet-B has 83 layers including 53 parameter layers, and SubEnsNet-C has 91 layers including 58 parameter layers. 
EnsembleNet also requires to configure a group of hyper-parameters for model initialization. For each basic DNN of EnsembleNet, the number of filters in the convolutions and the number of recurrent units are adjusted to be consistent with the number of learning features.
% Since our DNNs are insensitive to other hyper-parameters, such as kernel size and max-pooling strides, they can be tuned in a reasonable manner. 
In the training phase, we employ sparse categorical cross-entropy as loss function for performing error calculation that is used in the back-propagation; and root mean square propagation (RMSprop) algorithm \cite{tieleman2012lecture} as optimizer to minimize errors and accelerate gradient descent as well as convergence rate. Here, the learning rate is set to 0.001. In addition, as a rule of thumb, we configure the hyper-parameters in the correlation analyzer around their default values, where the number of components of PCA is set to 2 and the number of neighbors of both LOFs is set to 25.

\subsection{Evaluation Metrics}

Similar to other ML based designs, we use six metrics to evaluate the performance of the designs:

\begin{itemize}
    \item Accuracy (ACC): Number of correct predictions/Total number of predictions.
    \item Detection rate (DR): TP/(TP+FN), where TP is the number of attacks correctly categorized and FN is the number of attacks incorrectly classified as normal network traffic.
    \item False alarm rate (FAR): FP/(FP+TN), where FP is the number of actual normal traffic misclassified as attacks and TN is the number of normal network traffic correctly categorized.
    \item Precision: TP/(TP+FP).
    \item $F_1$ score: (2$\times$precision$\times$DR)/(precision+DR).
    \item Processing time (Time): Training time+testing time. 
\end{itemize}

\begin{figure*}[t]
    \centering
    \includegraphics[width=.98\linewidth]{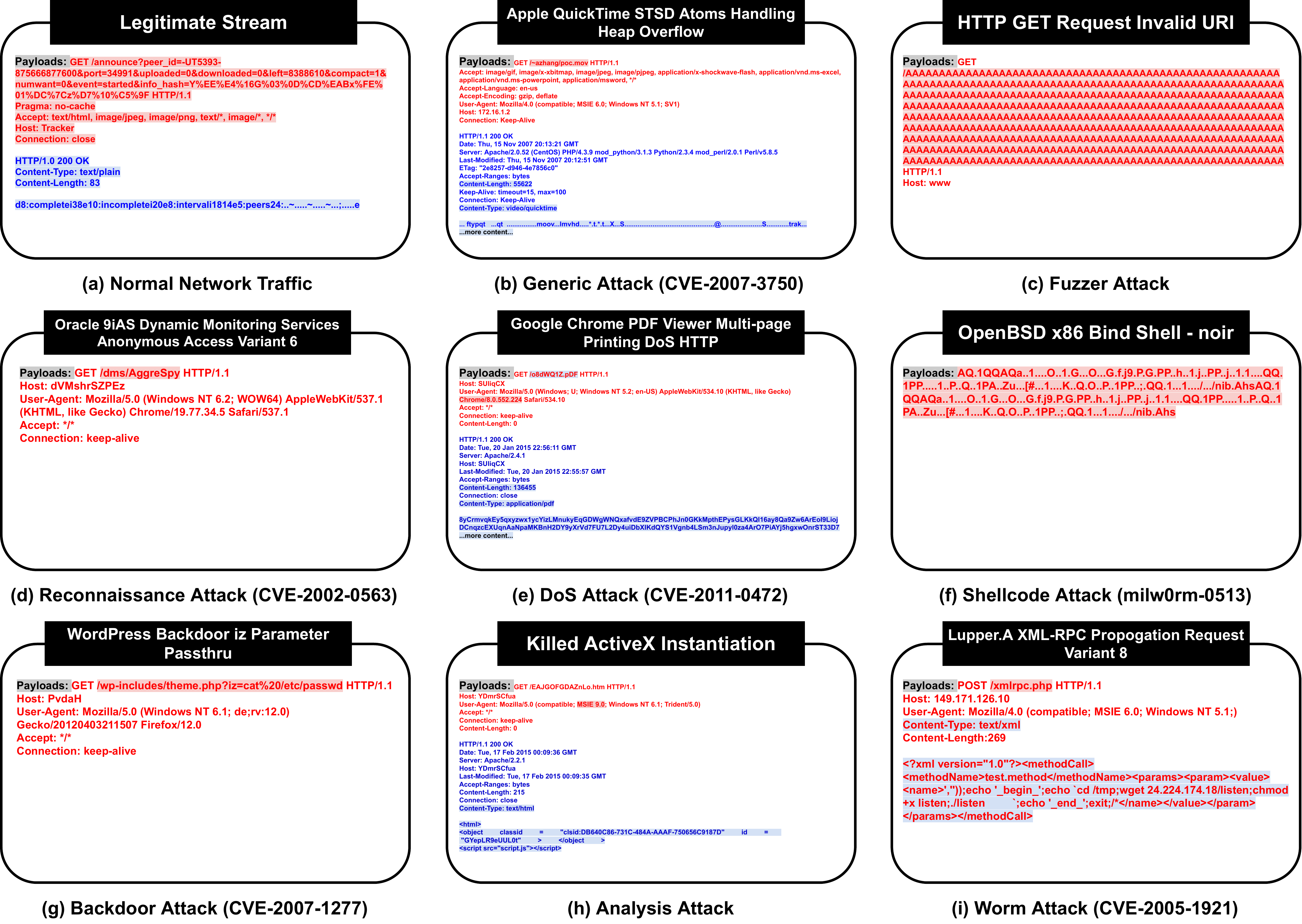}
    \caption{Comparison of the Payloads of the Normal Network Traffic and Different Attacks}
    \label{fig:comparison_attacks}
    \vspace{-1em}
\end{figure*}

\subsection{DarkHunter Performance}

\subsubsection{Overall Detection Performance}

Table \ref{tab:ensemble_nets} shows the testing performance and the processing time of SubEnsNet-A, SubEnsNet-B, SubEnsNet-C, EnsembleNet and DarkHunter. From the table, we can see that EnsembleNet outperforms all its sub-classifiers by achieving higher ACC on processing both multi-class and binary classification tasks, higher DR, higher precision, higher $F_1$ score and lower FAR, which testifies the effectiveness of proposed greedy majority voting algorithm. The algorithm only incurs around 0.05s (718.88s - 718.83s), which is ignorable. We can also find that the high sensitivity (due to high DR), high specificity (due to low FAR) and high dissimilarity of each sub-classifier of EnsembleNet contribute to the overall decision-making, enabling EnsembleNet to have a good performance. By using the correlation analyzer, the detection performance can be further improved, as shown in the last row of the table. DarkHunter can achieve the accuracy of 80.18\% on multi-class classification and 90.81\% on binary classification; and the detection rate, precision, $F_1$ score can reach to 98.13\%, 86.58\%, 91.99\%, respectively, while false alarm rate being reduced to 17.71\%.
The experimental results reflect the effectiveness of correlation analyzer in terms of further identifying intrusions and filtering out false positives. The correlation analyzer takes about extra 1.98s (720.86s - 718.88s), which is also ignorable considering the overall performance gain.

\subsubsection{Original Packet Stream Backtracking}

We randomly select 10\% of the records from the testing set. For each record, we obtain the packet stream generated by our flow tracing method and check whether the stream matches the real traffic flow of the record. There are 100\% matches, which confirms the effectiveness of our original packet stream backtracking strategy. 

In addition to the example given in Fig.~\ref{fig:interpretability}, here we include more examples of payloads, as shown in Fig.~\ref{fig:comparison_attacks},  to reveal the typical behaviour of other types of attacks detected by the neural network detection engine. For each attack, we also provide a brief description of the attack nature and a short summary of representative attack features.

\textbf{Normal Network Traffic.} Fig.~\ref{fig:comparison_attacks}(a) shows the payloads within a legitimate TCP stream: a normal payload of the HTTP request packet (marked in pink) and a normal payload of the HTTP response packet (marked in blue).

\textbf{Generic Attack.} Fig. \ref{fig:comparison_attacks}(b) shows the payloads within a TCP stream about a Generic attack, which is \textit{`Apple QuickTime STSD Atoms Handling Heap Overflow \cite{generic_attack}'  (CVE-2007-3750 \cite{cve_2007_3750}).} Apple QuickTime before 7.3 exists the heap-based buffer overflow vulnerability, which is due to boundary errors when processing Sample Table Sample Descriptor (STSD) atoms in a movie file. The hacker exploits the flaw to trick target users into opening a QuickTime movie file (marked in pink) with crafted STSD atoms (marked in blue), eventually leading to arbitrary code execution.

\textbf{Fuzzer Attack.} Fig. \ref{fig:comparison_attacks}(c) shows the payloads within a TCP stream about a Fuzzer attack, which is \textit{`HTTP GET Request Invalid URI \cite{fuzzer_attack}'.} The hacker continuously sends a series of HTTP GET requests with non-existent URLs (marked in pink) to the same destination address and destination port to analyze the response information to find and exploit potentially hackable vulnerabilities.

\textbf{Reconnaissance Attack.} Fig. \ref{fig:comparison_attacks}(d) shows the payloads within a TCP stream about a Reconnaissance attack, which is \textit{`Oracle 9iAS Dynamic Monitoring Services Anonymous Access Variant 6 \cite{recon_attack}' (CVE-2002-0563 \cite{cve_2002_0563}).} There is a default configuration flaw in the Oracle 9i Application Server version 1.0.2.x. The hacker exploits the vulnerability by accessing sensitive services anonymously without authentication, including Dynamic Monitoring Services such as servlet/DMSDump and DMS/AggreSpy (marked in pink).

\renewcommand{\baselinestretch}{1.1}
\begin{table}[t]
\caption{Testing Performance of Using DarkHunter for the Normal and Each Attack on UNSW-NB15 Testbed}
\begin{center}
\begin{tabular}{c||c|c|c|c|c}
\hline
\textbf{Category} & \textbf{ACC \%} & \textbf{DR \%} & \textbf{FAR \%} & \textbf{Precision \%} & \textbf{$F_1$} \\
\hline
\hline
Normal & 90.81 & 82.29 & 1.87 &	97.42  & 89.22 \\
\hline
Generic & 99.99 & 99.98 & 0.00 & 100.00 & 99.99\\
\hline
Exploit & 97.61 & 99.74 & 3.01 & 90.58 & 94.94\\
\hline
Fuzzer & 85.61 & 76.14 & 13.53 & 33.94 & 46.95\\
\hline
Recon & 99.22 & 99.91 & 0.84 & 90.41 & 94.92\\
\hline
DoS & 99.97 & 92.78 & 0.01 & 97.83 & 95.24\\
\hline
Shellcode & 99.70 & 97.30 & 0.29 & 46.75 & 63.16\\
\hline
Analysis & 98.41 & 0.00 & 1.59 & 0.00 & 0.00\\
\hline
Backdoor & 100.00 & 100.00 & 0.00 & 100.00 & 100.00\\
\hline
Worm & 100.00 & 100.00 & 0.00 & 75.00 & 85.71\\
\hline
\end{tabular}
\label{tab:each_attack}
\end{center}
\end{table}

\renewcommand{\baselinestretch}{1}

\renewcommand{\baselinestretch}{1.1}
\begin{table*}[t]
\caption{Testing Performance of Using Different Anomaly-Based Network Intrusion Detection Systems on UNSW-NB15}
\begin{center}
\begin{tabular}{c|c||c|c|c|c|c|c|c}
\hline

\multirow{2}{*}{\textbf{Type}} & \multirow{2}{*}{\textbf{IDS}} & \multicolumn{2}{c|}{\textbf{ACC \%}} & \multirow{2}{*}{\textbf{DR \%}} & \multirow{2}{*}{\textbf{FAR \%}} & \multirow{2}{*}{\textbf{Precision \%}} & \multirow{2}{*}{\textbf{$F_1$ Score}} & \multirow{2}{*}{\textbf{Time/s}}\\

\cline{3-4} 

& & \textbf{Multi-class} & \textbf{Binary} & & & & &\\
\hline
\hline

\multirow{4}{*}{Traditional ML} & AdaBoost & 52.29 & 74.30  & 92.56 & 48.06 & 70.23 & 79.86 & 71.95\\
\cline{2-9} 
& NB & 53.06 & 74.60 & 89.57 & 43.74 & 71.50 & 79.52 & 4.03\\
\cline{2-9}
& SVM & 54.51 & 63.89  & 69.40 & 42.85 & 66.49 & 67.91 & 2588.38\\
\cline{2-9}
& RF & 56.10 & 76.89 & 89.50 & 38.56 & 73.98 & 81.00 & 13.16\\
\hline

\multirow{6}{*}{Advanced DL} & LSTM & 68.88 & 84.78 & 94.68 & 27.35 & 80.92 & 87.26 & 326.59\\
\cline{2-9}
& ConvNet & 69.01 & 83.27 & 97.75 & 34.47 & 77.65 & 86.55 & 216.29 \\
\cline{2-9}
& MLP & 71.47 & 86.29 & 97.56 & 27.51 & 81.29 & 88.69 & 234.99\\
\cline{2-9} 
& Densely-ResNet & 72.92 & 85.64 & 95.34 & 26.24 & 81.66 & 87.97 & 1257.38\\
\cline{2-9}
& DualNet & 75.79 & 87.57 & 98.10 & 25.33 & 82.59 & 89.68 & 891.15\\
\cline{2-9}
& \textbf{DarkHunter} & \textbf{80.18} & \textbf{90.81} & \textbf{98.13} & \textbf{17.71} & \textbf{86.58} & \textbf{91.99} & \textbf{720.86}\\
\hline
\end{tabular}
\label{tab:comparison_results}
\end{center}
\end{table*}
\renewcommand{\baselinestretch}{1}

\textbf{DoS Attack.}
Fig.~\ref{fig:comparison_attacks}(e) shows the payloads within a TCP stream about a DoS attack, which is \textit{`Google Chrome PDF Viewer Multi-page Printing DoS HTTP \cite{dos_attack}' (CVE-2011-0472 \cite{cve_2011_0472}).} Google Chrome before 8.0.552.237 (marked in pink) has a vulnerability that can be triggered when a user prints a multi-page PDF document. The hacker launches the denial-of-service attack via the document (marked in blue), which would lead to an application crash or other unspecified impacts.

\textbf{Shellcode Attack.} Fig.~\ref{fig:comparison_attacks}(f) shows the payloads within a UDP stream about a Shellcode attack, which is \textit{`OpenBSD x86 Bind Shell -- noir \cite{shellcode_attack}' (milw0rm-0513).} The hacker transmits a block of shellcode (marked in pink) over a UDP socket to control the compromised machine.

\textbf{Backdoor Attack.} Fig.~\ref{fig:comparison_attacks}(g) shows the payloads within a TCP stream about a Backdoor attack, which is \textit{`WordPress Backdoor iz Parameter Passthru \cite{backdoor_attack}' (CVE-2007-1277 \cite{cve_2007_1277}).} During February and March 2007, WordPress 2.1.1 downloaded from several official distribution sites that included an externally introduced malicious backdoor. The hacker exploits the backdoor by executing arbitrary operating system commands via an untrusted passthru function call in the iz parameter to the wp-includes/theme.php (marked in pink).

\textbf{Analysis Attack.} Fig. \ref{fig:comparison_attacks}(h) shows the payloads within a TCP stream about an Analysis attack, which is \textit{`Killed ActiveX Instantiation \cite{analysis_attack}'.} The hacker sends a series of HTML pages that instantiate Microsoft ActiveX controls (marked in blue) to the same destination address and destination port, where the controls have set the kill bit through SPs or patches issued by Microsoft. These class identifiers (CLSIDs) are harmful if instantiated via Microsoft Internet Explorer (MSIE) (marked in pink), which can cause either command execution or memory corruption.

\textbf{Worm Attack.} Fig. \ref{fig:comparison_attacks}(i) shows the payloads within a TCP stream about a Worm attack, which is \textit{`Lupper.A XML-RPC Propogation Request Variant 8 \cite{worm_attack}' (CVE-2005-1921 \cite{cve_2005_1921}).}  Eval injection vulnerability in XML-RPC For PHP 1.1 and earlier version (marked in pink), as applied in WordPress, phpWebSite and other products. The Lupper.A worm exploits the bug to infect the system by executing a block of crafted PHP code via an XML file (marked in blue).

As can be observed through these examples and the example in Section \ref{section_4}, \textit{request target (URL)}, \textit{user agent}, \textit{content type} and \textit{message body} are strong features of payloads within a stream,  presenting the most valuable attack-related information that can be used for rapid attack identification, performing counter-attack measures and forensic analysis.
Furthermore, \textit{content length} is a weak feature that contributes to attack recognition and unknown threat perception.
It is worth noting that we have to be careful when the \textit{Post} method appears. The reason is that the Post pushes the data to the server, which could be a piece of crafted shellcodes.

\subsubsection{Detection Capability for Each Category}

DarkHunter can not only identify whether a flow is normal or abnormal but also determine its specific attack type if it is abnormal.
Table \ref{tab:each_attack} shows the testing performance of DarkHunter on different detection categories. From the table, we can see that DarkHunter performs well on the detection of normal traffic and most types of attacks -- with high ACC, high DR, high precision, high $F_1$ score, and low FAR, especially for Reconnaissance and DoS attacks (which is the benefit to early discovery of sophisticated threats such as APTs and DDoS attacks).
The exception is for the Fuzzer, Shellcode and Analysis attacks, which are discussed below.

% 60% standard: 
% Fuzzers: FAR, precision, f1
% Shellcode: precision
% Analysis: DR, precision, f1

For the Fuzzer attacks, the detection has a low precision and a high false alarm rate.
The main reason is that legitimate users may accidentally make typos when requesting valid URLs, thus confusing the classifier. 
For the Shellcode attacks, the precision is also low.
The most possible reason is that more than half of Shellcode records used in our evaluation are UDP-based, but the features used for learning are more TCP-specific \cite{moustafa2015unsw}, 
making learning not sufficient for such attacks and hence limiting its detection capability on this attack type.

The low performance on Analysis attacks (i.e. low DR, low precision, and low $F_1$ score) may due to two reasons.
One is that Analysis is to listen to and analyze network communications to capture basic cyber information. Its behavior can also be observed in the normal traffic, making it hard to distinguish the Analysis from the normal. For example, commands such as $whoami$ and $ipconfig$ can come from Analysis attacks but also can come from legitimate users. The Analysis attack is not a direct attack. It is rather considered as anomalous behavior that may lead to a real attack.
Another reason for the poor detection performance on the Analysis attack is that there are only around 1.04\% Analysis records applied in the evaluation, which is an imbalance learning problem that often results in poor generalization performance.

\subsubsection{Comparative Study}

To further evaluate the generalization performance of DarkHunter, we compare DarkHunter with a set of state-of-the-art ML-based IDSs. All designs have been discussed in detail in Section \ref{section_2}. Table \ref{tab:comparison_results} shows the testing performance of these IDSs. As can be observed from the table, the traditional ML methods accomplish high DR at the cost of high FAR.

The DL-based designs outperform the traditional ML methods with higher ACC on both multi-class and binary classification tasks, higher DR, higher precision, higher $F_1$ score and lower FAR.
Importantly, among these IDSs, the proposed defense mechanism DarkHunter presents the best overall performance; It has the highest ACC, highest DR, highest precision and highest $F_1$ score while maintaining the lowest FAR.
Although DarkHunter takes longer processing time than some of existing models (e.g. NB and ConvNet), its detection performance is far superior to those models. Among the DL models, Densely-ResNet and DualNet show comparative performance but their processing times are considerably higher than DarkHunter's. DarkHunter is more cost efficient. To sum up, the comparison results demonstrate that DarkHunter has a high generalization capability and the effectiveness of DarkHunter for network intrusion detection.

\section{Conclusion} \label{section_6}

In this paper, we have proposed a deep ensemble network based IDS, DarkHunter, for network intrusion detection. DarkHunter offers three design characteristics: 1) it combines both supervised learning and unsupervised learning, where unsupervised learning is for detection performance enhancement; 2) its detection engine is constructed with the deep network classifiers and their decisions are integrated with a greedy majority voting scheme; 3) it can generate user-friendly output alerts that are prioritized based on the threat severity level and are human understandable. 

We evaluate DarkHunter and compare it with a set of typical existing ML-based designs. Our experiment results on UNSW-NB15 dataset demonstrate that our design approach is effective and DarkHunter outperforms those ML-based designs. Among all the designs examined in our experiments, DarkHunter shows the highest capability of attack recognition while maintaining the lowest false alarm rate. 

It must be pointed out for DarkHunter to be used for real-time processing, an online dataset of sufficient size should be maintained for PCA and LOF to be effective, which will be further investigated in the future.

\section{Acknowledgement}

We thank Brian Udugama of University of New South Wales and Peilun Wu of Sangfor Technologies Inc., for their constructive feedback on the work presented in this paper.

\bibliographystyle{./bibliography/IEEEtran}
\bibliography{./bibliography/IEEEabrv,./bibliography/IEEEexample}

\end{document}